\documentclass{aa}
\usepackage{graphicx}
\usepackage{times}

\begin{document}

\title{The magnetic field along the jets of NGC\,4258\thanks{Based on
          observations with the 100-m telescope of the MPIfR
          (Max-Planck-Institut f\"{u}r Radioastronomie) at Effelsberg} }

\subtitle{as deduced from high frequency radio observations}

\author{M. Krause
  \and
   A. L\"ohr
   }

\offprints{M. Krause, email: mkrause@mpifr-bonn.mpg.de}
%\mail{mkrause@mpifr-bonn.mpg.de}

\institute{Max-Planck-Institut f\"ur Radioastronomie, Auf dem H\"ugel 69,
  53121 Bonn, Germany
}

\date{Received 6 August 2003 / Accepted 10 February 2004}

\abstract{
We present 2\farcs4 resolution, high sensitivity radio continuum
observations of the nearby spiral galaxy  NGC\,4258 in total intensity
and linear polarization obtained with the Very Large Array at
$\lambda$3.6~cm (8.44~GHz). \\
The radio emission along the northern jet and the center of the galaxy
is polarized and allows investigation of the magnetic field. Assuming
energy-equipartition between the magnetic field and the relativistic
particles and distinguishing between (1) a relativistic electron-proton
jet and (2) a relativistic electron-positron jet, we obtain average
magnetic field strengths of about (1) 310~$\mu$G and (2) 90~$\mu$G. The
rotation measure is determined to range from $-$400 to $-$800~rad/m$^{2}$
in the northern jet. Correcting the observed E-vectors of polarized
intensity for Faraday rotation, the magnetic field along the jet turns
out to be orientated mainly along the jet axis. An observed tilt with
respect to the jet axis may indicate also a toroidal magnetic field
component or a slightly helical magnetic field around the northern jet.
}

%\thesaurus{11(11.01.2; 11.10.1; 11.13.2; 11.19.1;NGC 4258)}

%\titlerunning{ }
%\authorrunning{ }

\maketitle

\section{Introduction}

The nearby galaxy NGC\,4258 (M\,106) is a bright SAB(s)bc spiral
(de Vaucouleurs et al.\ \cite{vauc}) at a distance of 7.2~Mpc
(Herrnstein et al.\ \cite{herr99}). It seems to possess a small bright
nucleus with a highly excited emission line spectrum (Burbidge et
al.\ \cite{bur}) and has been classified as a weakly active
Seyfert~2-type galaxy.

Most striking are the two so-called `anomalous arms', which are not
visible in the optical and were first detected in H$\alpha$ by
Court\`{e}s \& Cruvellier (\cite{cour}) in the inner region of the
galaxy. Van der Kruit et al. (\cite{kruit}) detected these anomalous
arms in the radio range where they extend out to the optical periphery.
Spectral index studies indicate that their radio emission is of
non-thermal origin (de Bruyn\ \cite{bruyn}; van Albada \cite{alb}).

The anomalous arms of NGC\,4258 have been extensively discussed in
terms of ejection of matter from the nucleus. The detection of a
water-maser (Claussen et al.\ \cite{claus}; Henkel et al.\ \cite{henk})
and an accretion disk around a supermassive central object (Miyoshi et
al.\ \cite{miyo}; Herrnstein et al.\ \cite{herr}), and the fact that
inner anomalous arms are orientated parallel to the rotation axis of
the accretion disk and can be traced even on subparsec scale
(Herrnstein et al.\ \cite{her}) indicate that the anomalous arms indeed
are jets. Whereas in the inner regions the anomalous arms clearly reveal
their jet character, many of the features at a greater distance from the
nucleus (e.g. their bifurcation) remain unexplained.

The three-dimensional geometry of the galaxy and its jets has been
discussed for a long time (cf. e.g. van Albada \& van der Hulst\
\cite{albhulst}; Hummel et al.\ \cite{hum}). The detection of an
accretion disk revealed directly for the first time that the central
part of the galaxy has a significant tilt with respect to the galactic
disk: the accretion disk itself has an inclination angle of $83\degr$
and a position angle (p.a.) of $86\degr$ (Miyoshi et al.\ \cite{miyo}),
i.e. it is oriented nearly east-west. The p.a. of the galactic disk,
however, is $150\degr$, which is nearly north-south, and its
inclination is $72\degr$ (van Albada \cite{albada}). Thus, the plane of
the galactic disk and the plane of the accretion disk have a
significant angle to each other. As the jets emerge perpendicular to
the accretion disk, they have to pass the galactic disk at a rather
small angle. Due to the projection of the whole system with respect to
the Earth, the inner jet direction is almost parallel to the major axis
of the galactic disk.

To further investigate the jet geometry and especially the
magnetic field along the jets we obtained data with the Very Large
Array (VLA)\footnote{The VLA is a facility of the National Radio Astronomy
Observatory. The NRAO is operated by Associated Universities, Inc.,
under contract with the National Science Foundation.} at
$\lambda$3.6~cm and compared these with already published but
reprocessed VLA data by Hummel et al. (\cite{hum}) at $\lambda$6.2~cm
(4.86~GHz) and $\lambda$20~cm (1.49~GHz) and with observations at
$\lambda$2.8~cm (10.55~GHz) made with the Effelsberg 100-m telescope.

The observations and data reduction procedures are described in
Sect.~2. In Sect.~3 we present the results and examine the measurements
of the linearly polarized emission in terms of Faraday rotation,
magnetic field strength and direction. The discussion of the results
and the summary follow in Sect.~4 and Sect.~5, respectively.

\section{Data acquisition and reduction}\label{obs}

\subsection{Observations with the VLA}

We observed the radio continuum emission from NGC\,4258 with the Very
Large Array (VLA) in its C-configuration at 8.44~GHz ($\lambda$3.6~cm)
in March 1996 for 14 hours in total and linearly polarized intensity.
The observations were done with two independent IFs, each with a
bandwidth of 50~MHz and separated by 50~MHz. The two IFs were combined
afterwards. The resulting observational parameters are given in
Table~\ref{tab1}. For a detailed description of the VLA see Thompson et
al. (\cite{tom}) and Napier et al. (\cite{nap}).

\begin{table}[h!]
\begin{center}
\caption{Observational parameters of the VLA measurements. The r.m.s.
noise values are given for total intensity (I) and polarized intensity
(I$_\mathrm{P}$). Further explanations are given in the text below.}
\label{tab1}
\begin{tabular}{l|c|c|c|c} \hline\hline
 & \multicolumn{3}{c|}{$\lambda$3.6~cm} &  $\lambda$6.2~cm   \\ \hline
Central frequency $$[GHz] & \multicolumn{3}{c|}{8.4399} & 4.8851 \\
Bandwidth [MHz] &  \multicolumn{3}{c|}{2$\times$50} & 2$\times$50   \\
Epoch & \multicolumn{3}{c|}{3/1996} &  7/1983  \\
Time on-source [hr] & \multicolumn{3}{c|}{14}  & 6  \\
Array & \multicolumn{3}{c|}{C}  & C  \\
Phase calibrator & \multicolumn{3}{c|}{1216+487}   & 1216+487  \\
Robust & 0 & 4 & 4 & 5 \\
Beam [arcsec] & 2.4 & 3.3 & 14 & 14 \\
r.m.s. (I) [$\mu$Jy/b.a.] & 8  & 8 & 70 & 60  \\
r.m.s. (I$_\mathrm{P}$) [$\mu$Jy/b.a.] & -- & 7 & 30  & 55   \\ \hline
\end{tabular}
\end{center}
\end{table}

The data reduction was performed using the Astronomical Image
Processing System (AIPS) of the NRAO.

The flux density scale was calibrated by observing 3C\,138 and 3C\,286
and is based on the Baars et al. (\cite{bars}) scale. We assumed flux
densities of 2.52~Jy for 3C\,138 and 5.21~Jy for 3C\,286.
We used an angle of $-12\degr$ for the polarized emission of 3C\,138
and $-33\degr$ for that of 3C\,286. The phase  calibrator 1216+487,
whose position is known with an accuracy of $\leq 0\farcs1$, was also
used to correct for the instrumental polarization.

The edited and calibrated visibility data were cleaned interactively,
self-calibrated and Fourier transformed to obtain maps of the
Stokes parameters I, U and Q. We produced uniformely weighted maps
(with ROBUST=0) to obtain the best resolution and sensitivity
compromise with a HPBW of $2\farcs2 \times 2\farcs4$. Maps even more
sensitive to weak extended structures and the polarized emission were
made with natural weights (ROBUST=4) at the expense of a slightly lower
HPBW of $2\farcs9 \times 3\farcs3$. We used the zero flux correction
provided by AIPS as a first-order correction for the missing
large-scale flux. The U- and Q-maps were combined to obtain maps of the
linearly polarized intensity I$_\mathrm{P}$ (corrected for the positive
zero level offset) and the position angles of the polarized emission
(the E-vectors). Values of the respective noise levels are summarized in
Table~\ref{tab1}, the full resolution map is shown in Fig.~\ref{fig1}
and Fig.~\ref{fig2}.

For a comparison we used and reproduced already published VLA data by
Hummel et al. (\cite{hum}) at 4.86~GHz ($\lambda$6.2~cm) and 1.49~GHz
($\lambda$20~cm). The observational parameters at $\lambda$6.2~cm are
listed in Table~\ref{tab1}. The entire reduction was done in the same
way as described above. The full resolution maps obtained at
$\lambda$6.2~cm and $\lambda$20~cm have HBPWs of $14\arcsec$; the
$\lambda$6.2~cm map is shown in Fig.~\ref{vla6cm}. For a
comparison we also smoothed the naturally weighted maps at
$\lambda$3.6~cm to $14\arcsec$ HPBW by folding with Gaussian beams in
two steps with the AIPS task CONVL. Noise level values of the
$\lambda$6.2~cm and the smoothed $\lambda$3.6~cm maps are also
included in Table~\ref{tab1}. The data at $\lambda$20~cm were of
little use for the comparison because the polarized intensity was
only detected at large distances from the center.

\subsection{Observations with the Effelsberg 100-m telescope}

NGC~4258 was observed with the Effelsberg 100-m telescope
at 10.55~GHz ($\lambda$2.8~cm) in July 1995. The multi-beam receiver
has 4 horns and 16 channels. Each horn is equipped with two total-power
amplifiers and an IF polarimeter. The bandwidth is 300~MHz, the system
noise temperature about 50~K and the resolution is $69\arcsec$ HPBW.

For pointing and focusing we observed regularly the sources 3C\,48 and
3C\,286. The calibration was done with 3C\,286 according to the flux
values of Baars et al. (\cite{bars}). The observing procedure is
essentially the same as described by Klein \& Emerson (\cite{kle}). We
obtained 28 coverages of NGC~4258 which led, after restoration (Emerson
et al. \cite{emer}) and combination (Emerson \& Gr\"ave \cite{emergr}),
to an r.m.s. noise in the final maps of $900~\mu$Jy/beam in total
power and $300~\mu$Jy/beam in linear polarization.

\section{Results}

\subsection{Total intensity}

\begin{figure}[htb]
\includegraphics[bb = 36 107 576 652,width=8.8cm,clip=]{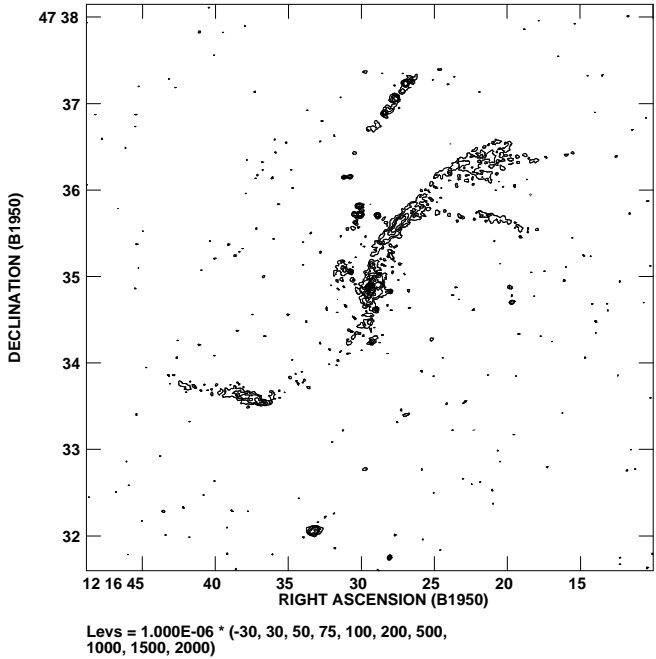}
\caption{The 8.44~GHz high resolution (HPBW~=~$2\farcs2 \times
2\farcs4$) map of total intensity as obtained with the VLA
C-configuration. The r.m.s. noise is $8~\mu$Jy/b.a.}
\label{fig1}
\end{figure}

\begin{figure}[htb]
\includegraphics[bb = 46 108 561 642,width=8.8cm,clip=]{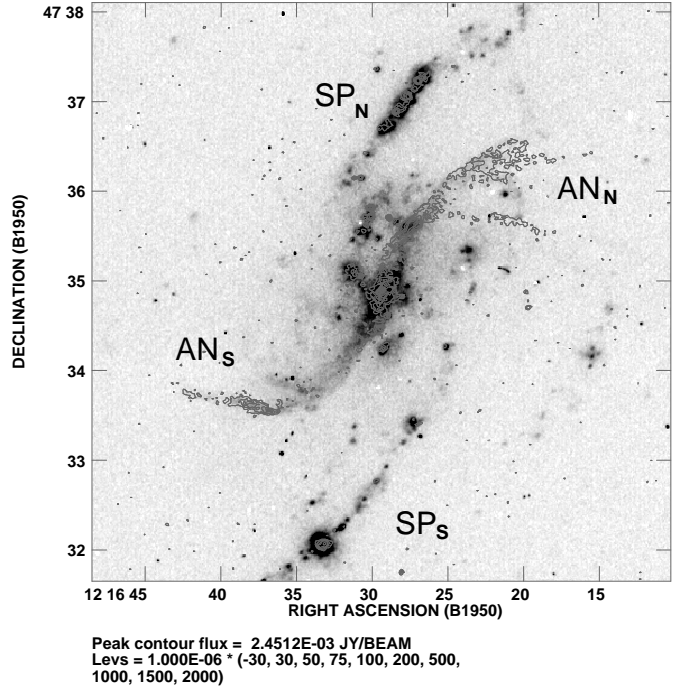}
\caption{The 8.44~GHz high resolution (HPBW~=~$2\farcs2 \times
2\farcs4$) map in total intensity as obtained with the VLA
C-configuration superimposed on the H$\alpha$ map of the galaxy
observed at the Hoher List Observatory of the University of Bonn (by
courtesy of N. Neininger). The normal spiral arms are marked as SP$_\mathrm{N}$
and SP$_\mathrm{S}$, the anomalous arms as AN$_\mathrm{N}$ and AN$_\mathrm{S}$.}
\label{fig2}
\end{figure}

Figure~\ref{fig1} shows the full resolution map of NGC\,4258 in total
intensity at 8.44~GHz as obtained with the VLA in its C-configuration.
The HPBW is $2\farcs2 \times 2\farcs4$ and the inner region of the
galaxy is well resolved. The S-shaped feature of the anomalous arms is
clearly visible (marked as AN$_\mathrm{S}$ and AN$_\mathrm{N}$ in
Fig.~\ref{fig2}) with the southern jet being fainter than the
northern one. The jet direction in the central part is approximately north
to south, whereas it changes to about east-west orientation in the outer
parts. The northern arm bifurcates at least two times (as first
detected by van Albada \& van der Hulst\ \cite{albhulst}).

Maximum radiation is emitted from the galactic center. The normal
spiral arms of the disk are also visible to the north and
south (marked as SP$_\mathrm{N}$ and SP$_\mathrm{S}$ in Fig.~\ref{fig2}).
Along them, three very bright regions can be observed at
$\rm\delta_{50}=47\degr 32\arcmin 10\arcsec\ (\rm{SP}_\mathrm{S})$,
$\rm\delta_{50}=47\degr 35\arcmin 45\arcsec\ (\rm{SP}_\mathrm{N}1)$ and
$\rm\delta_{50}=47\degr 37\arcmin 20\arcsec\ (\rm{SP}_\mathrm{N}2)$,
where the intensity is $200~\mu$Jy/b.a. ($\rm{SP}_\mathrm{S}$), $500~\mu$Jy/b.a.
(SP$_\mathrm{N}$1) and
$200~\mu$Jy/b.a. (SP$_\mathrm{N}$2).

We observe a ridge of high intensity of
$100~\mu$Jy/b.a. along the northern jet. The southern jet shows a maximum
brightness of $100~\mu$Jy/b.a. at large distance from the center. Between the
center and this region the southern jet is very weak. The northern region of
high brightness ($80~\mu$Jy/b.a.) at $\alpha_{50}=12\fh16\fm28.9$
and $\delta_{50}=47\degr 35\arcmin 43\arcsec$ (marked as `A' in
Fig.~\ref{fig3}) has been interpreted as a radio hot spot of the northern
jet by Cecil et al. (\cite{ce}). The same source has been classified as
a supernova remnant candidate by Hyman et al. (\cite{hym}), noted as
source 7 therein.

The corresponding hot spot in the southern jet (marked as `B' in Fig.~\ref{fig3})
has been reported to be located $24\arcsec$ south of the nucleus at
$\alpha_{50}=12\fh16\fm29.6$ and $\delta_{50}=47\degr 34\arcmin
30\arcsec$ (Cecil et al. (\cite{ce})). In this region we also detected a local maximum in brightness
of $50~\mu$Jy/b.a. which coincides with the change of the jet
direction as indicated by the solid line in Fig.~\ref{fig3}.

The total flux of the central source was measured by fitting a
two-dimensional Gaussian to the central emission in the map with highest
resolution (Fig.~\ref{fig1}). It is derived to be $2.45\pm 0.09$~mJy.

In Fig.~\ref{fig2} the contour plot of the total intensity is
superimposed on an H$\alpha$ map of the galaxy (observed at the
Hoher List Observatory of the University of Bonn, (courtesy of N.
Neininger). In H$\alpha$
the central region as well as the normal spiral arms to the north
and south show the highest intensity. The superposition assigns
clearly which part of the radio emission belongs to the spiral arms
and which belongs to the jets. The clumps of strong emission along the
spiral arms mentioned above are well correlated with the regions of
highest intensity in H$\alpha$. They are most propably star-forming
regions as they emit unpolarized thermal radiation, whereas the radio
emission along the anomalous arm is polarized and of nonthermal origin
(cf. e.g. Hummel et al.\ \cite{hum}; Hyman et al.\ \cite{hym}).

\begin{figure}[htb]
\includegraphics[bb = 72 123 567 672,angle=270,width=8.8cm,clip=]{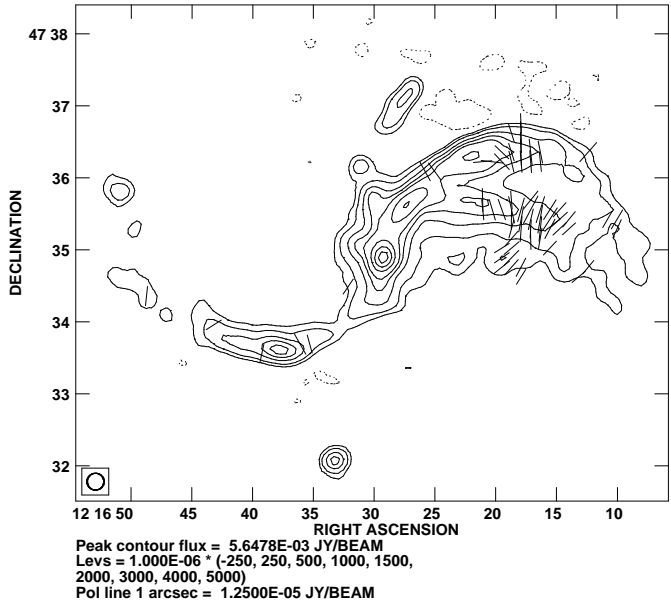}
\caption{The 4.86~GHz map in total intensity with HPBW~=~14\arcsec.
The r.m.s. noise is $60~\mu$Jy/b.a. The length of the E-vectors is
proportional to the linearly polarized intensity.}
\label{vla6cm}
\end{figure}

\begin{figure}[htb]
\includegraphics[bb = 36 100 576 660,width=8.8cm,clip=]{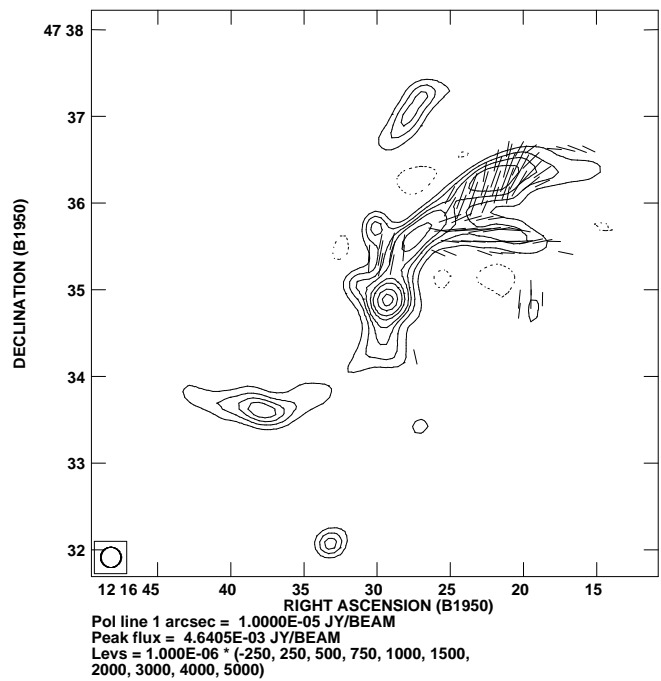}
\caption{The 8.44~GHz map of the total intensity at HPBW~=~14\arcsec.
The r.m.s. noise is $70~\mu$Jy/b.a. The length of the E-vectors is
proportional to the linearly polarized intensity.}
\label{fig4}
\end{figure}

Figure~\ref{vla6cm} shows the observations of NGC~4258 at
4.86~GHz ($\lambda$6.2~cm) as observed with the VLA in its C-configuration.
The contours give the total intensity, the length of the E-vectors is
proportional to the linearly polarized intensity. The r.m.s. noise
values are given in Table~\ref{tab1} and the HPBW is 14\arcsec.
To compare the $\lambda$3.6~cm map with the $\lambda$6.2~cm
map we smoothed the $\lambda$3.6~cm to the resolution of $14\arcsec$
HPBW. This map is shown in Fig.~\ref{fig4}. The shape of the
total intensity is very similar at both wavelengths.

\begin{figure}[htb]
\includegraphics[bb = 46 107 567 658,width=8.8cm,clip=]{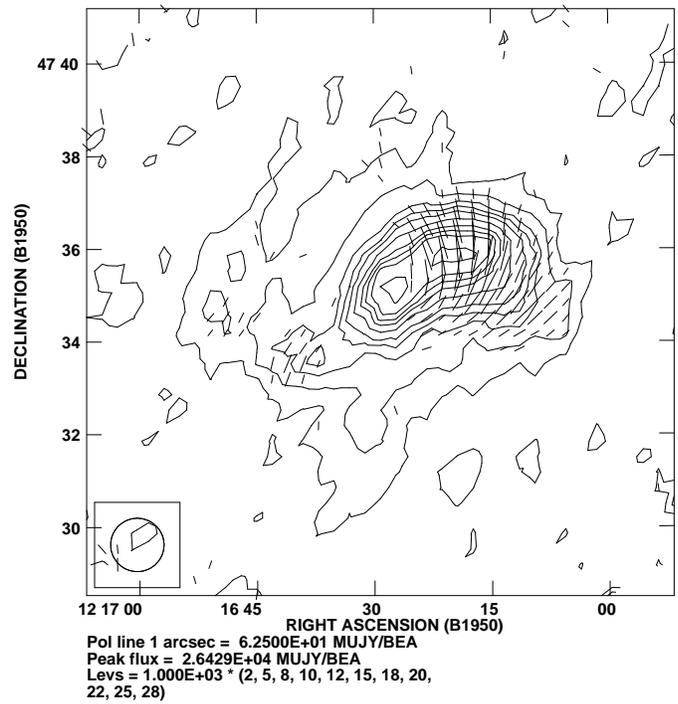}
\caption{The 10.55~GHz map of NGC4258 in total intensity (contours)
with HPBW$=69\arcsec$. The r.m.s. noise is $900~\mu$Jy/b.a. The length
of the E-vectors is proportional to the linearly polarized intensity.}
\label{eff}
\end{figure}

The observations at $\lambda$2.8~cm made with the Effelsberg 100-m
telescope are presented in Fig.~\ref{eff}. The HPBW is 69\arcsec, hence
the distribution is much smoother and the bifurcation in the western
arm is barely resolved. The normal spiral arms, notably the southern
one, are indicated.

\subsection{Jet geometry}

\begin{figure}[htb]
\includegraphics[bb = 77 159 536 636,width=8.8cm,clip=]{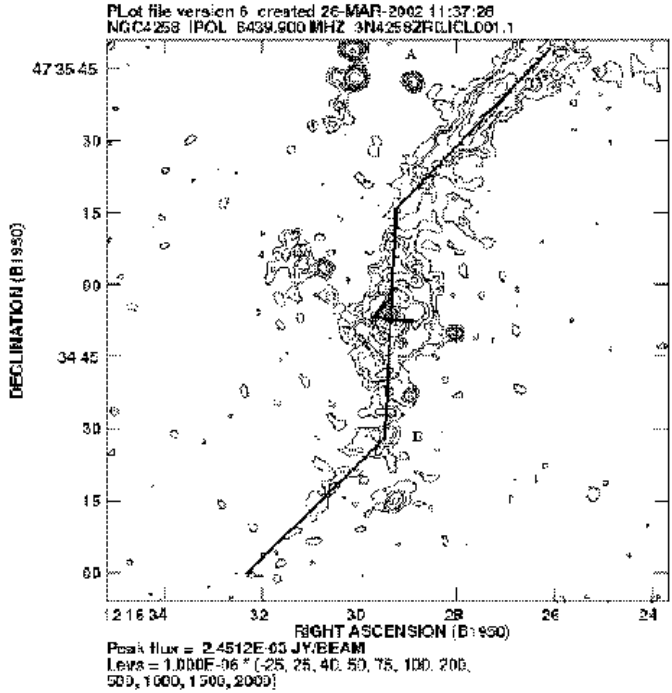}
\caption{The 8.44~GHz high resolution (HPBW~=~$2\farcs2 \times
2\farcs4$) map detail of the central region in total intensity. The r.m.s.
noise is $8~\mu$Jy/b.a. The location of the circumnuclear
accretion disk and the observed geometry of the jets (solid lines) as well
as the supposed hot spots A and B are indicated.}
\label{fig3}
\end{figure}

To investigate the jet geometry and in particular the outflow
direction from the nucleus, Fig.~\ref{fig3} shows the central region in
detail up to $60\arcsec$ away from the nucleus. The location of the circumnuclear accretion disk and the observed direction of the ejected
matter are indicated. From our observations we determined the
position angle of the jet in the central region to be $\rm p.a.=-3\degr
\pm 1\degr$. This is in full agreement with the value given
by Cecil et al. (\cite{ce}) from their radio observations and along
the projected spin axis of the accretion disk as determined by Miyoshi
et al. (\cite{miyo}).

At larger distances from the nucleus the jets change their direction
from a north-south orientation to northwestern to southeastern
as visible on larger scales (cf. also Fig.~\ref{fig1}). This
change is not smooth but can be described by symmetric kinks on
both sides at a projected distance of $24\arcsec$ from the galactic
center to $\rm p.a.= -43\degr \pm 1\degr$.

As described by Martin et al. (\cite{mart}) both anomalous
arms bifurcate as seen in H$\alpha$. The projected nuclear distance of
the bifurcation is, however, non-symmetric: the southern jet splits
at about $84\arcsec$ and the northern jet at about $62\arcsec$
from the nucleus, both to $\rm p.a.= 71\degr$. The northern jet reveals
two more bifurcations, at $104\arcsec$ and at about $125\arcsec$
(marginally detected) projected distances from the nucleus.

\subsection{Polarized intensity}

\begin{figure}[htb]
\includegraphics[bb = 36 109 576 652,width=8.8cm,clip=]{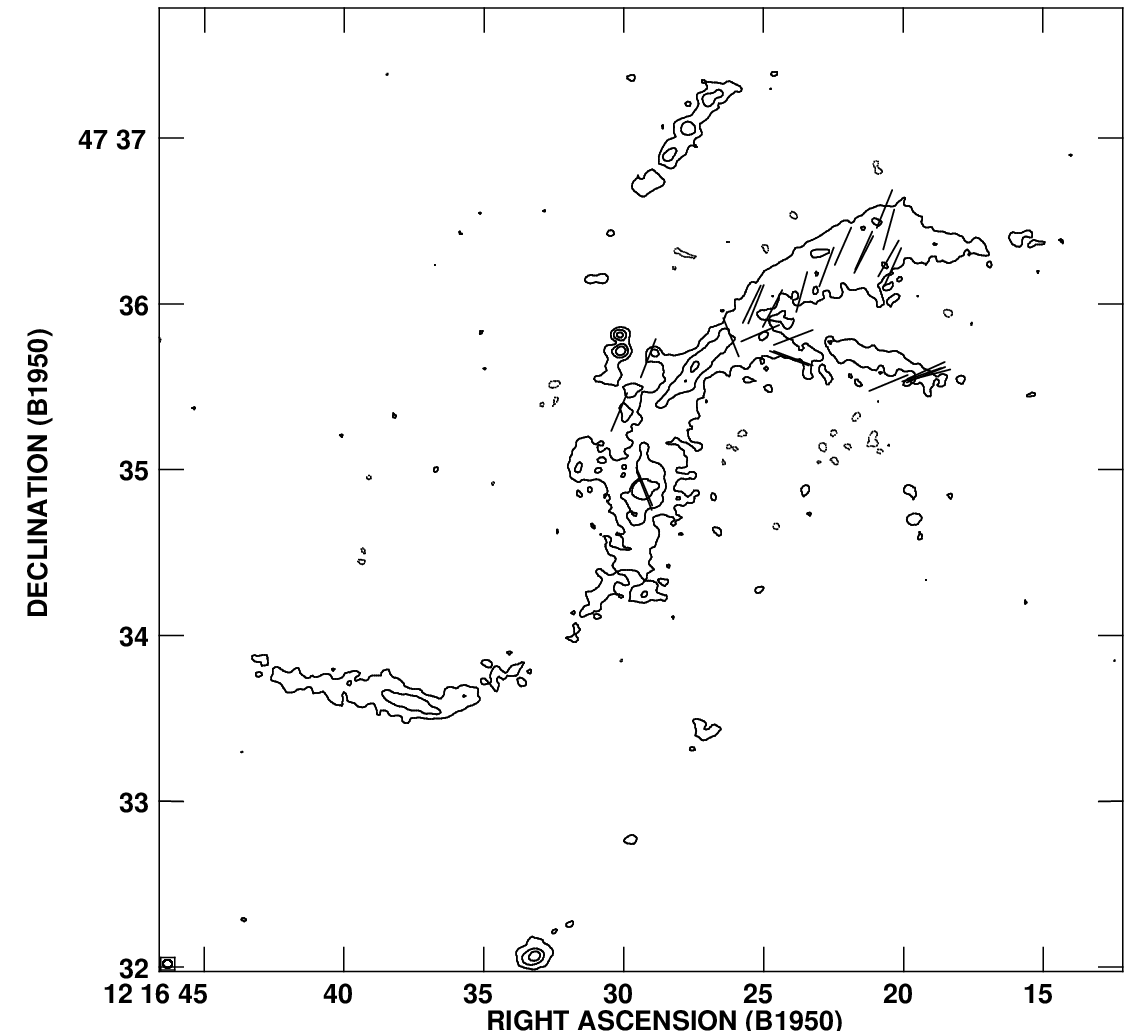}
\caption{The 8.44~GHz map of the total intensity with HPBW~=~$2\farcs9
\times 3\farcs3$. The r.m.s. noise is $8~\mu$Jy/b.a. The length of the
E-vectors is proportional to the linearly polarized intensity.}
\label{fig6}
\end{figure}

In addition to the total intensity we obtained the linearly polarized
intensity at a resolution of $2\farcs9 \times 3\farcs3$ HPBW. This is
the first time that extended linear polarization has been detected in
NGC\,4258 with arcsec resolution. In Fig.~\ref{fig6} the E-vectors of
the linearly polarized emission are superimposed on the total
intensity maps at 8.44~GHz. The length of the vectors is proportional
to the polarized intensity.

As can be seen, polarized intensity has been detected exclusively along
the jets. The highest polarized emission occurs at the center and along
the northern jet, there mainly to the northwest of the bifurcation.
The map shown in Fig.~\ref{fig6} will be used later to calculate the
magnetic field strength (Sect.~3.6).

The degree of linear polarization (polarized intensity/total
intensity) along the jets is high and lies in the range between
35\% and 65\%. In the center the degree of linear polarization is below
2\% and may be partly instrumental. A high degree of polarization of
about 55\% was detected at the presumed hot spot A, whereas
no polarization was detected at the position of source B. This
will be discussed further in Sect.~4.

In the smoothed map (Fig.~\ref{fig4}) the highest polarized emission
was detected along the jets and is particularly strong along the
northern jet. The polarization percentage varies between 20\% and 45\%
there. These values are lower than those of the higher resolution map
(Fig.~\ref{fig6}) which indicates that the polarized intensity is
patchy and no longer resolved at $14\arcsec$ HPBW. The smoothed map
also reveals several small regions of low percentage polarization
(20\%) in the southern jet.

A comparison of the $\lambda$3.6~cm map (Fig.~\ref{fig4}) to the
$\lambda$6.2~cm map (Fig.~\ref{vla6cm}) shows very different
orientation of the E-vectors in both maps, indicating strong Faraday
rotation.

\subsection{Faraday rotation measure and depolarization}

The observed electric vectors are rotated by Faraday effects. The
amount of the Faraday rotation can be determined by calculating the
rotation measure RM between different wavelengths. Correction of the
observed electric vectors according to these RMs and rotation by
$90\degr$ leads to the {\em intrinsic} direction of the magnetic field
in the sky plane. The RM value itself depends on the strength of the
magnetic field component parallel to the line of sight, its sign
indicates the direction of this parallel field component.

Although also we had $\lambda$20~cm data, we could not use
them for the calculation because at this wavelength and resolution
polarized intensity was only detected at large distances from the
center. At $\lambda$3.6~cm and $\lambda$6.2~cm the polarized intensity
was detected closer to the nucleus and along the jets. Thus the
polarized regions detected do not coincide with those at $\lambda$20~cm.

We determined the RM between $\lambda$3.6~cm and $\lambda$6.2~cm.
The calculated RM varies between 400 and $\rm 800~rad/m^2$. The $\rm
n\pi$ ambiguity (i.e. the RM value that corresponds to a Faraday
rotation of $\rm n 180\degr$ and hence is undistinguishable by
observations at only two wavelengths) between these two wavelengths is
as high as $\rm 1216~rad/m^2$ for n=1. A substraction of one $180\degr$
rotation (n=-1) from these values leads to RM between $-$800 and $\rm
-400~rad/m^2$ which is in absolute value not distinguishable from the
values above. Futhermore, a rotation of the observed vectors at
$\lambda$3.6~cm by e.g. $\rm -400~rad/m^2$ or $\rm 816~rad/m^2$
(which corresponds to the $\pi$ ambiguity) leads to an angle
difference of $88\degr$, hence about perpendicular to each other.

\begin{figure}[htb]
\includegraphics[bb = 52 109 569 660,width=8.8cm,clip=]{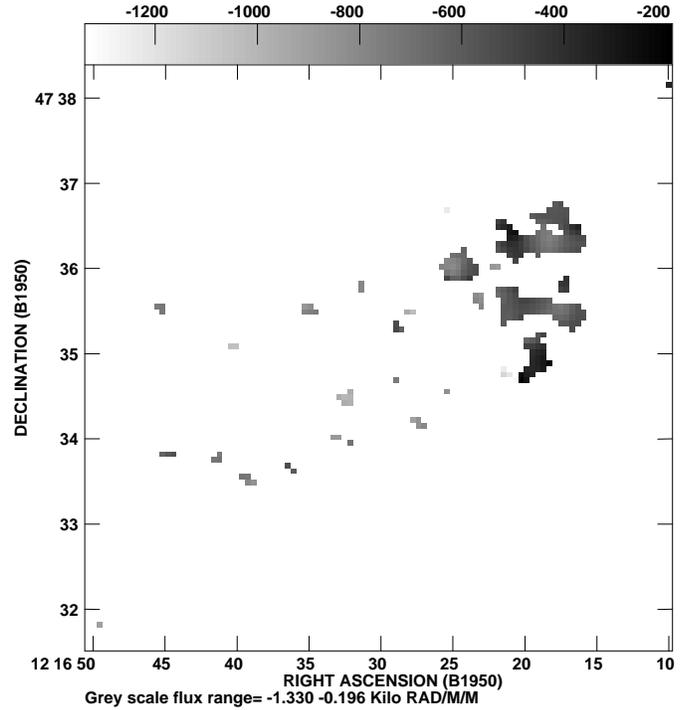}
\caption{The rotation measure RM between $\lambda$3.6~cm and
$\lambda$6.2~cm (HPBW~=~14\arcsec).}
\label{fig7}
\end{figure}

We tried to solve the RM ambiguity using the $\lambda$2.8~cm
observations. Therefore we smoothed the $\lambda$3.6~cm map
(Fig~\ref{fig4}) out to 69\arcsec~HPBW, the angular resolution of
the $\lambda$2.8~cm map (Fig.~\ref{eff}), and determined the RM between
these two wavelengths. The $\rm n \pi$ ambiguity between these
wavelengths is as large as about $\rm 7200~rad/m^2$. The vectors
between $\lambda$2.8~cm and $\lambda$3.6~cm rotate {\em clockwise} by
about 40--$80\degr$ which correspond to negative RM between about
$-$1500 and $\rm -3000~rad/m^2$. We conclude that the vectors rotate
further {\rm clockwise} towards the E-vectors observed at
$\lambda$6.2~cm. This corresponds to negative RM values. Hence we
consider the values for n=1 between $\lambda$3.6~cm and $\lambda$6.2~cm
as more probable. The corresponding RM ranges from $-$800 to $\rm
-400~rad/m^2$ and is presented graphically in Fig.~\ref{fig7}.

As the galactic foreground rotation measure is negligible towards the
direction of this galaxy (Krause et al.\ \cite{mari}) we conclude that
the observed RM is excusively associated with NGC\,4258.

To verify the consisitency of the derived values we estimate
the depolarization $DP$ between $\lambda$3.6~cm and $\lambda$6.2~cm
using
\[ DP=\frac{\textrm{polarization  percentage}_\mathrm{\lambda
    6.2\,cm}}{\textrm{polarization percentage}_\mathrm{\lambda
    3.6\,cm}}\ . \]
>From our observations we derive $DP \approx 0.3$ in the northern jet.
A uniform magnetic
field {\em in} the jet leads to differential Faraday depolarization
(Burn\ \cite{burn}; Sokoloff et al.\ \cite{depol})
\[DP = \left| \frac{\sin(2 RM \lambda^2)}{2 RM \lambda^2} \right|
 \ . \]
Using this formula we obtain $|\rm RM|$ of 600--650~rad/m$^2$ for
$DP\approx 0.3$, which is in perfect agreement with the values calculated
above for the RM from the rotation of the observed E-vectors in that area.
The rotation of the electric vectors between $\lambda$3.6~cm and
$\lambda$6.2~cm that belongs to $\rm |RM|$ of 600~rad/m$^2$ is already
more than $90\degr$. {\em If} the Faraday rotation (and depolarization)
takes place inside the emitting region we have to conclude that the
Faraday depth of the jet at $\lambda$6.2~cm is smaller than the jet
itself, hence we do not see the full jet in linear polarization as part
of its linearly polarized emission cancels with that from other parts to
form unpolarized emission.

However, the observed RM may also be due to a layer of higher thermal
density but fewer relativistic particles around the jet that does not
contribute to the polarized emission as proposed by Bicknell et al.
(\cite {bick}). This layer would act as a foreground screen and
rotate the observed vectors without depolarizing the emission.

The observed depolarization of the emission can also be explained by
beam depolarization and Faraday dispersion (Burn\ \cite{burn}; Sokoloff
et al.\ \cite{depol}) which is expected to occur in the emitting
region, especially in the case of a toroidal magnetic field (see
Sect.~3.5) that is not resolved by the beam size.

With the observed large values for the RM we expect strong
depolarization at $\lambda$20~cm in all scenarios mentioned above.
This can explain the lack of linearly polarized emission in the inner
part of NGC\,4258 as observed by Hummel et al. (\cite{hum}).

\subsection{Magnetic field direction}

\begin{figure}[htb]
\includegraphics[bb = 49 112 566 654,width=8.8cm,clip=]{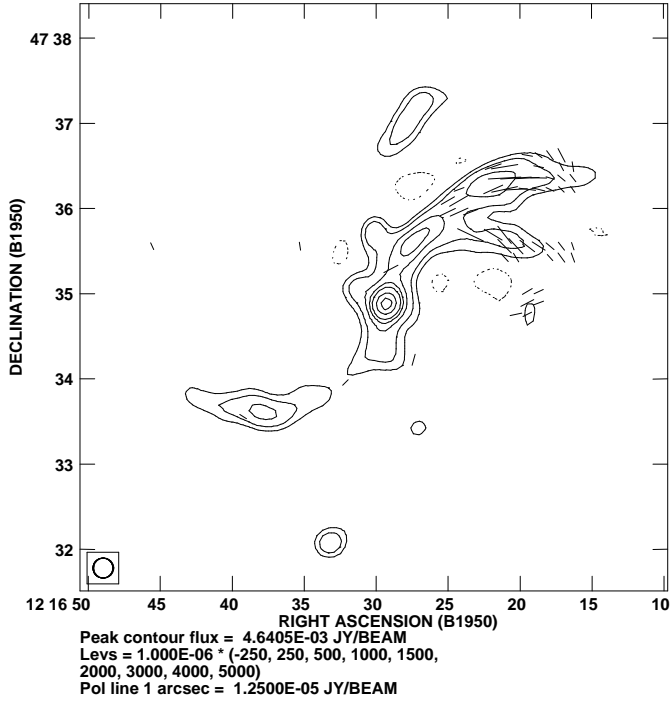}
\caption{The 8.44~GHz map of the total intensity (HPBW~=~14\arcsec) as
shown in Fig~\ref{fig4}. The r.m.s. noise is
$70~\mu$Jy/b.a. The B-vectors show the magnetic field direction with
their length proportional to the polarized intensity.}
\label{fig5}
\end{figure}

To obtain the direction of the intrinsic magnetic field, we corrected
the E-vectors (Fig.~\ref{fig4}) for Faraday rotation with the help of
the RM (Fig.~\ref{fig7}). The resulting magnetic field is shown in
Fig.~\ref{fig5} superimposed on the smoothed $\lambda$3.6~cm map.
The vectors generally follow the jet direction, also along the main
bifurcation of the northern jet. They are, however, somewhat tilted
with respect to the jet direction, especially in the outermost part of
the northern bifurcation of the northern jet and along the southern
part of this bifurcation. We infer that the magnetic field is mainly
poloidal along the jet axis with a (weaker) toroidal component
especially in the outer part of the northern bifurcation and along the
southern bifurcation. This can be a superposition of a poloidal and
(weaker) toroidal magnetic field in different layers around the jet
axis or is consistent with a slightly helical magnetic field
around the jet axis.

\subsection{Magnetic field strength}

To calculate the magnetic field strength we used the polarized data at
$\rm HPBW\approx 3\arcsec$ (Fig.~\ref{fig6}). As the polarized
emission is not homogeneously distributed but varies strongly along the
jets, we only calculate the magnetic field in the regions of strongest
polarized emission, which are: region I northeast of the bifurcartion,
region II northwest of the bifurcation and region III just south of the
bifurcation.

For the nonthermal spectral index along the jets we used the latest
value derived by Hyman et al. (\cite{hym}) which is
$\alpha_\mathrm{n}=-0.65\pm0.10$. This value is in good agreement
with the average spectral index in jets of $\alpha_\mathrm{n}=-0.7\pm
0.1$ and also with the previously derived spectral index between
$\lambda$6.2~cm and $\lambda$20~cm by Hummel et al. (\cite{hum}) of
$\alpha_\mathrm{n}=-0.6\pm0.1$.

The jet is assumed to be cylindrical. For the line of sight $L$
through the jets we assume $L=50$~pc as we do not resolve the jet
width with our linear resolution of 90~pc. Based on our previous
results in Sect.~3.5 we assume a magnetic field mainly along the
jet axis. As the inclination of the jet we considered values between
$5\degr$ and $45\degr$ (as argued in Sect.~4).

We calculate the magnetic field strengths assuming energy equipartition
between the magnetic field and the relativistic particles. The
calculations consider two different models concerning the composition of
the ejected jet-matter in active galactic nuclei:
\begin{enumerate}
\item{The jets contain relativistic electrons and protons similar to
  the cosmic rays observed near the Earth. According to Ginzburg \&
  Syrovatskij (\cite{gin}) the electron energy density in the relevant
  energy interval is then 1\% of that of the proton energy density and
  thus negligible. The K-factor used in this case is 100 (Krause et
  al.\ \cite{mari}; Beck\ \cite{beck}).}
\item{The plasma jets contain relativistic electrons and positrons
  indistinguishable by observation. Then the total energy density is
  given by both sorts of particles. The K-factor used in this case is 1.}
\end{enumerate}

Table~\ref{tab2} shows the magnetic field strengths in the three
regions along the northern jet for an electron-proton jet and
Table~\ref{tab3} shows the results for an electron-positron jet.

\begin{table}[h!]
\begin{center}
\caption{Magnetic field strengths for an electron-proton jet. Region
I is south of the northern bifurcation, II is the southern
part of the N bifurcation; III is the northern part of the N bifurcation.}
\label{tab2}
\begin{tabular}{c|ccc} \hline
Region & $B_\mathrm{total}$  & $B_\mathrm{uniform}$ & $ B_\mathrm{random}$ \\
   & [$\mu$G] & [$\mu$G] & [$\mu$G]  \\ \hline \hline
I    & 325 $\pm$ 16 &  210 $\pm$ 20 & 250 $\pm$ 25 \\
II   & 310 $\pm$ 15 &  250 $\pm$ 25 & 185 $\pm$ 18 \\
III  & 300 $\pm$ 15 &  270 $\pm$ 27 & 130 $\pm$ 13  \\ \hline
\end{tabular}
\end{center}
\end{table}

\begin{table}[h!]
\begin{center}
\caption{Magnetic field strengths for an electron-positron jet.
Regions I, II, and III are chosen as in Table~\ref{tab2}.}
\label{tab3}
\begin{tabular}{c|ccc} \hline
Region & $B_\mathrm{total}$ &$B_\mathrm{uniform}$ &$B_\mathrm{random}$ \\
   &  [$\mu$G] & [$\mu$G] & [$\mu$G] \\  \hline \hline
I    &  92 $\pm$ 5 & 60 $\pm$ 6 & 70 $\pm$ 7 \\
II   &  88 $\pm$ 4 & 70 $\pm$ 7 & 53 $\pm$ 5 \\
III  &  85 $\pm$ 4 & 75 $\pm$ 8 & 37 $\pm$ 4 \\ \hline
\end{tabular}
\end{center}
\end{table}

\section{Discussion}

The large-scale radio and H$\alpha$ features of NGC\,4258 could only
be interpreted as jets (Falcke \& Biermann\ \cite{fal}; Yuan et al.\
\cite{yuan}) after the accretion disk around a super-massive central
object had been discovered from observations of water-maser emission (Claussen
et al.\ \cite{claus}; Henkel et al.\ \cite{henk}). Previously they were
called `anomalous arms' and explained in terms of ejected matter that
interacts strongly with the disk gas and hence is compressed (van der
Kruit et al.\ \cite{kruit}) or as jet-like outflows (Sanders et al.\
\cite{sand}).

The detection of a water-maser implies a thin, rotating
Keplerian disk and is strong evidence for a black hole in the center
of NGC\,4258 (Miyoshi et al.\ \cite{miyo}). Jets require enormous
amounts of energy and it is generally assumed that they are produced by
accretion disks which rotate around black holes and eject matter along
their rotation axis.

The jets in NGC\,4258 are, however, extraordinary jets, as they are not
strongly collimated but show a diffuse structure and even bifurcation.
They are also strongly bent.

The three-dimensional geometry of this galaxy is somewhat unusual in
that the accretion and galactic disk are almost perpendicular
to each other (cf. Sect.~1). If the jets emerge almost perpendicular
to the nuclear disk, they have to pass the galactic disk and seem to
interact with it at least in the inner 4~kpc (cf. Krause et al.\
\cite{mari2}).

Our data are the first that are sensitive enough to detect the magnetic
field in the northern jet regions also in the inner $2\arcmin$ from
the galactic center. We determined the rotation measure and corrected
the polarization angles for Faraday rotation. The derived intrinsic
magnetic field orientation is mainly along the jet direction.

As the determined RMs are negative, the uniform magnetic field
component along the jet axis points away from us, hence it is directed
towards the nucleus in the northern jet. The few RM values that are
found in the southern jet are also negative and may indicate that the
magnetic field there is directed away from the nucleus.

The magnetic field in the northern jet however is also somewhat tilted
towards the jet direction, especially in the outermost part of the northern
bifurcation and along the southern part of the bifurcation. This can
indicate an additional (weaker) toroidal component that is either
located in different layers around the jet axis or may be due to a
partly helical magnetic field around the jet axis. This is found to be
typical of extragalactic jets (Begelmann et al.\ \cite{begel}).

The longitudinal field (along the jet axis) can either be in the inner
part of the jet near the jet axis, in the so-called {\em beam}, with a
toroidal field further away from the jet axis (e.g. Roland \& Hermsen\
\cite{rol}) or be part of a helical magnetic field (e.g. Lesch et al.\
\cite{lesch}) that may be amplified by dynamo action such as the screw
dynamo (Shukurov \& Sokoloff\ \cite{dyn}).

As $\mathrm{RM} = 0.81 \int n_\mathrm{e} B_\parallel dl$ (the sign
of RM is determined by the direction of $B_\parallel$), the observed
high RM values require a uniform magnetic field component $B_\parallel$
{\em parallel to the line of sight with uniform direction}.
It has been argued from the geometry of the accretion disk (see
Sect.~1) that the inner jet orientation is almost parallel to the major
axis of the galactic disk of NGC\,4258. A magnetic field along the jet
that is parallel to the major axis of the disk would not contribute to
a field component parallel to the line of sight, and hence not
cause Faraday rotation. However, even an angle of the jet to the line of
sight of only $5\degr$ yields $B_\parallel = 0.1~B_\mathrm{long}$,
where $B_\mathrm{long}$ is the field component along the jet axis.
Additionally, we observe several kinks in the northern jet as described
in Sect.~3.2. The first appeared at $24\arcsec$ projected
distance from the nucleus where the jet changes direction by as
much as $40\degr$ in the plane of the sky. We consider it highly
improbable that this change of direction happens only in the
plane of the sky. It will take place at an arbitrary angle to the
line of sight.

Let us assume that we have a kink of equal strength along the
line of sight. With the uniform magnetic field strength of about
$250~\mu$G we estimate $B_\parallel$ of about $160~\mu$G in the
case of an electron-proton jet. We calculate the thermal electron
density $n_\mathrm{e}$ using
\[ \mathrm{RM}=0.81 n_\mathrm{e} B_{\parallel} L. \]
With RM~=~650~rad/m$^2$, $B_{\parallel} =160~\mu$G, and $L=50$~pc we
derive $n_\mathrm{e} = 0.1~\mathrm{cm}^{-1}$. This value is about
three times higher than usual values for the intersteller medium. A
higher value is reasonable as both jets are visible in H$\alpha$.

If the jet consists of relativistic particles (electron-positron
jet), the expected rotation measure {\em in} the jet is RM~=~0 and the
Faraday rotation is produced in a cocoon around the jet (e.g.
Bicknell et al.\ \cite{bick}) with a correspondingly smaller line of
sight, hence an even higher thermal electron density and/or magnetic
field strength there. In this case the expected depolarization is
smaller (as described in Sect.~3.4.) but still compatible with our
observations.

Previous investigations on the magnetic field were only able
to determine the magnetic field orientation in the outermost parts of
the jets (outside
$r\simeq 2\arcmin$) (cf. Krause et al.\ \cite{mari}; Hummel et al.\
\cite{hum}). The magnetic field there was also derived to be {\em along}
the jets.

Concerning the hot spots detected by Cecil et al. (\cite{ce}), we can
confirm that we also detected high emission at both locations. Hot
spots are usually detected when the relativistic jet matter hits
intergalactic gas and undergoes strong interaction. The observed electric
vector at $\lambda$3.6~cm fits, however, into the general pattern of
the northern jet. As we have no high-resolution polarization
information at $\lambda$6.2~cm we cannot determine the RM and the {\em
intrinsic} magnetic field direction at the presumed hot spot A.

On the other hand, the observed high degree of linear polarization of
about 55\% as averaged over the whole source makes it rather
improbable that source A is a supernova remnant as has been proposed
by Hyman et al. (\cite{hym}).

\section{Conclusions}

We present interferometer data obtained with the VLA in its
C-configuration at $\lambda$3.6~cm (8.4399~GHz) in total power and
linear polarization. For comparison and to obtain quantities
like rotation measure and depolarization we also reprocessed VLA
data at $\lambda$6.2~cm (4.8851~GHz) and $\lambda$20~cm (1.4899~GHz)
that were previously published by Hummel et al. (\cite{hum}) and
observations at $\lambda$2.8~cm (10.55~GHz) made with the Effelsberg
100-m telescope.

The high resolution maps ($2\farcs4$ HPBW) at $\lambda$3.6~cm as
obtained with the C-configuration of the VLA are able for the first time
to resolve the jets in the central region of the galaxy. Detailed
inspection shows that they emerge from the galactic center along the
projected spin axis of the accretion disk as determined by Miyoshi et
al. (\cite{miyo}). At a distance $24\arcsec$ away from the center
they change direction towards the previously seen northwest to
southeast and bifurcate at $62\arcsec$ from the nucleus in
the northern jet and at $85\arcsec$ in the southern jet. The multiple
splitting of the northern jet is clearly visible at this resolution.

The polarized emission was detected exclusively along the jets and
allowed the calculation of the magnetic field strength in the inner
region of the northern jet. Energy equipartition considerations lead
to magnetic field strengths of $310\pm 15~\mu$G assuming a relativistic
electron-proton jet and $90\pm 5~\mu$G assuming an electron-positron
jet. The rotation measure could be determined between $\lambda$3.6~cm
and $\lambda$6.2~cm, at a linear resolution of $14\arcsec$ HPBW. It is
derived to be $-$400 to $-$800~rad/m$^2$ in the northern jet. Correcting
the observed E-vectors of polarized emission for Faraday rotation, the
magnetic field is mainly along the jet axis in the
central region and tends to become somewhat tilted with respect to the
jet direction in the outer part of the northern jet. This may be
consistent with a slightly helical magnetic field around the northern
jet or may indicate a superposition of a longitudinal magnetic field
near the jet axis and a toroidal magnetic field away from the axis.

\begin{acknowledgements}
We thank N. Neininger for providing the H$\alpha$ image. We
acknowledge fruitful discussions with H. Falcke and A. Shukurov
and are grateful for helpful comments by R. Perley.
\end{acknowledgements}

\end{document}